\documentclass[journal,twoside]{IEEEtran}

\ifCLASSINFOpdf
\usepackage[pdftex]{graphicx}

\else

\fi

\usepackage[cmex10]{amsmath}
\usepackage{amssymb}   
\usepackage{amsxtra}
\usepackage{amscd}
\usepackage{amsthm}
\usepackage{textcomp}
\usepackage{graphicx}  
\usepackage{color} 
\usepackage{MnSymbol}

\usepackage{flushend}

 \usepackage[table,xcdraw]{xcolor}

\usepackage{balance}

\usepackage{array}  

\usepackage{multirow}  

\usepackage{amsmath}
\usepackage{cite}

\usepackage{url}

\usepackage{color,soul} 
\soulregister\cite7
\soulregister\ref7
\soulregister\pageref7

\begin{document}
	%
	
	\title{{\huge Kirchhoff Meets Johnson: In Pursuit of Unconditionally Secure Communication }}

	
	%
	%
	%
	
	\author{Ertugrul~Basar,~\IEEEmembership{Fellow,~IEEE}
		\vspace*{0.35cm}
		\thanks{The author is with the Communications Research and Innovation Laboratory (CoreLab), Department of Electrical and Electronics Engineering, Ko\c{c} University, Sariyer 34450, Istanbul, Turkey. (e-mail: ebasar@ku.edu.tr).}
	}

\maketitle

\begin{abstract}
	
	Noise: an enemy to be dealt with and a major factor limiting communication system performance. However, what if there is gold in that garbage? In conventional engineering, our focus is primarily on eliminating, suppressing, combating, or even ignoring noise and its detrimental impacts. Conversely, could we exploit it similarly to biology, which utilizes noise-alike carrier signals to convey information? In this context, the utilization of noise, or noise-alike signals in general, has been put forward as a means to realize unconditionally secure communication systems in the future. In this tutorial article, we begin by tracing the origins of thermal noise-based communication and highlighting one of its significant applications for ensuring unconditionally secure networks: the Kirchhoff-law-Johnson-noise (KLJN) secure key exchange scheme. We then delve into the inherent challenges tied to secure communication and discuss the imperative need for physics-based key distribution schemes in pursuit of unconditional security. Concurrently, we provide a concise overview of quantum key distribution (QKD) schemes and draw comparisons with their KLJN-based counterparts. Finally, extending beyond wired communication loops, we explore the transmission of noise signals over-the-air and evaluate their potential for stealth and secure wireless communication systems. 
	
\end{abstract}
\begin{IEEEkeywords}
	Key distribution, Kirchhoff-law-Johnson-noise (KLJN) scheme, thermal noise communication (TherCom), thermal noise modulation (TherMod), quantum key distribution (QKD), unconditional security.  
\end{IEEEkeywords}

%
\IEEEpeerreviewmaketitle

\section{Introduction}

The 5G-Advanced standard (3GPP Release 18) is currently under progressive development. Its purpose is to enhance the performance of existing 5G networks with various advancements, including multiple-input multiple-output (MIMO) evolution, reduced capability user equipment services, artificial intelligence-empowered and data-driven networks, improved positioning, and reconfigurable intelligent surface-based systems. It is undeniable that the recent 5G standards have been developed with innovative physical layer (PHY) technologies in mind, such as millimeter-wave communications, massive MIMO systems, and multiple orthogonal frequency division multiplexing numerologies. These technologies are tailored to meet the diverse requirements of services and applications. With that said, despite providing a bold vision for future communication networks, our community has widely argued that 5G is not a revolutionary networking paradigm; instead, it is an evolution of 4G in terms of the adapted PHY solutions. As a result, active research efforts have already begun to uncover new communication paradigms for 6G wireless networks, anticipated for 2030 and beyond. These paradigms aim to offer not only high spectrum and energy efficiencies but also improved security at the PHY level, even in challenging environments \cite{Wang_2023}. Given the rise of potential 6G applications, secure communication will be more vital than ever, necessitating incorporating security considerations into the design of communication networks \cite{Arslan_2023}.

Within the context of revolutionary communication paradigms for future wired/wireless networks, using noise and noise-like signals presents a potential new paradigm that can achieve two primary objectives: unconditionally secure communication and low/zero-power communication. In the realm of classical electrical engineering, noise is viewed as an enemy to be dealt with, while we believe there might be gold in that garbage! In essence, rather than trying to suppress, eliminate, or disregard noise as is typically done in modern systems, we can exploit it, similar to how biology employs noise-like signals (e.g., the brain's use of noise-like signals as carriers of information), to unveil unconventional communication systems that offer unprecedented levels of secrecy. Specifically, for legacy communication systems, noise has long been identified as the primary obstacle to achieving high capacity, secrecy, reliability, and range. Contrary to this perspective, we pose the following question: can we leverage noise itself to establish reliable and unconditionally secure communication? Our aim is to explore answers to this intriguing question.

This review article presents the evolution of noise-driven communication systems across the past, present, and future. Despite its tutorial nature, this article makes noticeable contributions. We begin by emphasizing statistical-physical key exchange schemes that utilize Kirchhoff circuit loops and are triggered by the Johnson (thermal) noise. Within this framework, we delve into the fundamentals of noise-driven communication. Subsequently, we provide a historical perspective on the key distribution problem and ponder the necessity of physics-based key exchange schemes for secure communication. In this context, we also delve into the basics of quantum key distribution (QKD) schemes and provide a high-level comparison with noise-driven statistical-physical key exchange schemes. Finally, breaking the Kirchhoff loops, we explore wireless communication by evaluating thermal noise for secure and stealth communications.

Concerning the existing tutorials and survey papers on evolving technologies and security solutions for future networks, this article stands out uniquely due to its broad coverage of unconditional security, noise-driven communication, stealth communication, and physics-based key distribution schemes. The organization of the article is summarized as follows. Sections II and III present the concept of noise-driven communication and Johnson noise-based key distribution. Section IV gives a brief historical overview of secure communications and emphasizes the need for physics-based key distribution schemes. Sections V and VI present QKD protocols and their high-level comparison with classical statistical-physical (noise-driven) key exchange systems. Finally, noise modulation over wireless channels is covered in Section VII, along with conclusions in Section VIII.

\section{Roots of Thermal Noise Communication}
The origins of communication using modulated thermal noise date back to the early 2000s. Nevertheless, it still needs to be a fully perceived paradigm by our community despite its vast potential. Recently, this paradigm has been resurrected under the general theme of 
\textit{thermal noise communication (TherCom)} \cite{Basar_2023}. This article will try to further shed light on this promising paradigm for future wired and wireless networks.

The concept of zero-signal-power and stealth communication by means of thermal noise was first introduced in the pioneering work of Kish \cite{Kish_2005}. Here, an information bit is represented by a selection from two different impedance bandwidths (capacitors), creating two possible thermal noise spectra that the receiver observes. In other words, a unique form of thermal noise spectrum modulation (noise spectrum indexing) is envisioned. In this scenario, the considered transmitter does not emit signal energy to the channel; instead, it modulates the parameters of the thermal noise to convey information. Similarly, one can give examples such as mirror flashes in ancient Greek and Roman communications, the heliograph used in the 19th century, and communication by means of the reflected power introduced in the 1940s \cite{Stockman_1948}. It is worth noting that despite using inherent background noises, this concept is different from energy-free transmitters since the communicator devices at both sides of the link consume power for bit encoding and detection.

In the same study by Kish \cite{Kish_2005}, a wireless communication system operating in a line-of-sight (LOS)-dominated short communication link with two parabolic antennas is also envisioned. According to the information bits in this system, the transmitter switches between a resistor and an open (or short) circuit. One of the significant advantages of this method is its stealth nature since it might be very difficult, or even impossible, to discover the operation of a communication system using only background noises. However, despite its stealth nature, this noise-driven wireless system was not intended for unconditionally secure communication and might be subject to security breaches once its operation is discovered. For instance, a hidden eavesdropper can quickly determine the transmitted bit by assessing the resulting noise spectrum. This raises the question: can we use the thermal noise, either natural (background) or externally generated noises, for secure communication? Furthermore, we want this security to be at the level of quantum-based systems, that is, unconditional as guaranteed by the laws of physics.

Despite plain TherCom schemes not being able to provide unconditional security, a unique system utilizing thermal noise sources to obtain unconditional security is discussed in the following.

\begin{figure*}[!t]
	\begin{center}
		\includegraphics[width=1.6\columnwidth]{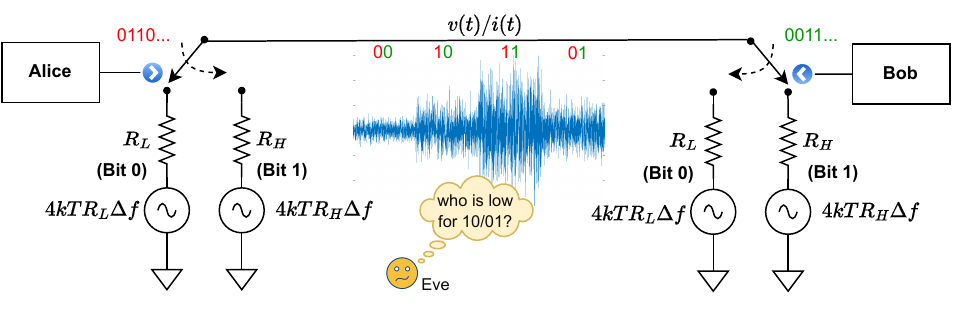}
		\vspace*{-0.3cm}\caption{KLJN secure bit (or key) exchange scheme. Three example thermal noise voltage variances are shown for $00$, $01/10$, and $11$ cases ($00\rightarrow$ low, $01/10\rightarrow$ intermediate, and $11\rightarrow$ high). $k$: Boltzmann’s constant, $T$: temperature in Kelvin degrees, $\Delta f$: bandwidth, $v(t)/i(t)$: noise voltage/current waveform.}\vspace*{-0.3cm}
		\label{fig:KLJN}
	\end{center}
\end{figure*}

\section{Kirchhoff-Law-Johnson-Noise Scheme}

Taking the noise spectrum modulation concept for zero-power communication one step further, the unconditionally secure \textit{Kirchhoff-law-Johnson-noise, (KLJN)} key exchange (generation) scheme was proposed in 2006 by exploiting the Kirchhoff's law and thermal noises of two pairs of resistors \cite{Kish_2006,Kish_2006_New}. This section presents a detailed overview of KLJN, followed by a discussion on KLJN bit detection.

\subsection{Fundamentals of KLJN}
In simple terms, KLJN is a statistical-physical system consisting of four resistors and related measurement as well as communication systems, which can generate and share an unconditionally secure binary key between two communicating parties (Alice and Bob), ensuring no information leak to an eavesdropper (Eve) under ideal conditions. In this setup, the current and voltage through the wire are random, ideally Johnson noise, and can be measured by anyone. The characteristic features of this noise are determined by the combination of resistors at each end, which ensures unconditional secrecy \cite{Kish_2016}. Furthermore, any active attack always introduces energy into the system that Alice and Bob can easily spot in the case of thermal equilibrium \cite{MIT_2012}. At this point, we observe the conceptual similarity with quantum-based key generation systems where measuring the quantum system also disturbs it.

One can ask the following question regarding the KLJN system: \textit{Compared to the bulletproof security provided by quantum weirdness, can a simple wire that is freely exposed to the public, with measurable voltage and current waveforms, offer unconditional security for eavesdroppers with unlimited measurement capabilities?} The answer is yes, thanks to the laws of physics (thermodynamics). In other words, like its quantum-based counterpart, violating the security of KLJN requires violating specific laws of physics.
	
	One of the reasons why our community has not fully understood the scheme of KLJN in the past 15 years is due to its multidisciplinary nature, which includes statistical physics, circuit noise theory, stochastic signals, and information-theoretic security. In the following subsection, we will discuss the basic technical operation principles of the KLJN scheme.
	
	\subsection{KLJN Operation}
	As shown in Fig. 1, for the operation of the KLJN system, the two available users (Alice and Bob) first perform resistor selection according to their randomly generated bits in each transmission interval. Then, they connect these selected resistors to a wire channel to create a Kirchhoff loop between themselves. Here, we assume that Alice and Bob are perfectly synchronized in time. For both users, bit $ 0 $ and bit $ 1 $ are represented by low- and high-valued resistors ($R_L$ and $R_H$), respectively. Here, Alice and Bob consider the measurements of the noise voltage $ (v(t)) $ and/or current $ (i(t))  $  waveforms on the wireline to decode their partner's bit. Specifically, considering Kirchhoff’s law with two parallel resistors, the power spectral density (or voltage variance per hertz of bandwidth) of the resulting noise voltage on the line, which is, in fact, a zero-mean Gaussian random process, is obtained as follows
	\begin{equation}
		S_v(f)=4kT \frac{R_A R_B}{R_A + R_B} \quad \text{V$^2$/Hz},
	\end{equation}
	which can alternatively be expressed in units of $\text{dBm/Hz}$ or watts/Hz considering a reference impedance. In (1), $R_A$ and $R_B$ represent the randomly selected resistors by Alice and Bob, respectively, that is, $R_A,R_B\in \left\lbrace R_L, R_H \right\rbrace $. Accordingly, for the four possible cases of selected Alice/Bob bits, namely $00,01,10,$ and $11$, where the first and second bits respectively stand for the selected bits by Alice and Bob, voltage waveform samples on the line will be Gaussian distributed random variables with different variances. The corresponding variance values of the noise samples are given as follows
	\begin{align}
		\sigma_{00}^2&= 4kT\frac{R_L R_L}{R_L+R_L} \Delta f=4kT\frac{R_L }{2} \Delta f \nonumber \\
		\sigma_{01}^2&=\sigma_{10}^2= 4kT\frac{R_L R_H}{R_L+R_H} \Delta f=4kTR_L\frac{ \alpha}{1+\alpha} \Delta f \nonumber \\
		\sigma_{11}^2&= 4kT\frac{R_H R_H}{R_H+R_H} \Delta f=4kTR_L\frac{ \alpha}{2} \Delta f,
		\label{eq:variances}
	\end{align}  
	where $\alpha$ is the ratio of resistances, that is, $R_H=\alpha R_L$. As illustrated in Fig. \ref{fig:KLJN}, the fluctuations on the line will depend on the selected resistance values at both sides of the system. Specifically, the fluctuations on the line will be high if both Alice and Both select the large resistance $R_H$. On the other hand, if both Alice and Bob select the small resistance $R_L$, the fluctuations will be small. What about the intermediate level? If one user (Alice or Bob) selects the large resistance while the other selects the small one, the noise variance on the line takes an intermediate value, which lies at the heart of the KLJN operation \cite{Cho_2005}. This intermediate noise variance value creates an indeterminacy for an eavesdropper (Eve), thus paving the way for unconditional security. Before providing more detailed discussions on this important topic, let us give a numerical example of the noise variances of different cases. Specifically, for the value of $\alpha=10$, noise variances on the line for the cases of intermediate and high fluctuations would be $1.8182$ and $10$ times greater than that of small fluctuations, respectively.
	
	An unconditionally secure bit exchange occurs when the bit values (or equivalently resistances) at the two ends of the KLJN system differ\footnote{Although a random key is established between Alice and Bob by the KLJN system, it has been widely regarded as a key exchanger in literature due to its unique architecture.}. For this case, an intermediate level is observed for the noise voltage variance (mean-square noise voltage), as shown in Fig. 1. Even though this intermediate level can be easily detected by Eve from a sufficient number of noise samples taken over the line, the specific contributions of Alice and Bob cannot be understood. This confusion on Alice's and Bob's resistances, created deliberately for Eve, ensures unconditional security at the level of quantum-based key distribution systems. In other words, the KLJN scheme consistently creates an indeterminacy for external observers, rendering the intermediate noise variance meaningless. In contrast, the cases of $ 00 $ and $ 11 $, respectively, provide the lowest and highest noise variances on the link. In this case, an eavesdropper can identify Alice/Bob bits from its measurements without a significant effort, corresponding to a non-secure bit exchange.
	
	As stated earlier, under ideal conditions, the cases of $ 01 $ and $ 10 $ produce equivalent noise variances for the noise samples. As a result, the specific locations of Alice/Bob bits cannot be determined from measurements taken on the line between the users. This unique property of KLJN provides a sort of powerful encryption and ensures absolute (unbreakable) security. For Alice and Bob, determining their partner's selected bit is straightforward for all four cases using voltage and/or current samples and the knowledge of their own bit. However, they have to take a sufficient number of noise samples for each transmission interval to reduce their bit errors. While we provide brief coverage of this issue in the following subsection, interested readers are referred to \cite{Basar_2023} for a comprehensive theoretical evaluation of KLJN's bit error probability (BEP) and its different types of detectors exploiting noise voltage and/or current samples.
	
	It is worth noting that the switching mechanism of the KLJN scheme among low and high resistances has conceptual similarities with the index modulation (IM) method \cite{IM_5G}, which embeds information in certain system entities such as antennas, subcarriers, and time slots. To be more specific, a KLJN communicator performs a unique form of IM by considering the indices of its available two resistors, or equivalently, for three possible noise voltage (or current) power spectral densities; however, it requires a no-wave solution for security. As in all IM systems, index selection does not consume power, and the primary task of the users is to extract the information embedded in the corresponding system entity through IM, which is the noise variance for the case of the KLJN scheme. In this context, selecting the number of noise samples per bit is critical, as discussed in the following subsection.
	
	While concluding this subsection here, Section VI discusses the state-of-the-art KLJN solutions and their security vulnerabilities.

	\subsection{Conquering the Noise: KLJN Detection}
	Noise: an enemy to be dealt with for achieving reliable and secure communication. Deviating from mainstream engineering, we can ask the following question: is it possible to use the noise itself, for instance, as biology does using noise-alike signals as the information carrier, for communication purposes? KLJN uses the noise itself for this purpose; as a result, we need to explore practical ways to extract the information embedded in this noise. In this subsection, we shed light on this issue.
	
	To determine their partner's bit in the KLJN system, Alice and Bob take samples from the thermal noise voltage (or current) on the wire for each bit duration $(T_b)$. The total number of ``uncorrelated" noise samples taken per bit, denoted by $N$, can be derived from the Wiener–Khinchin theorem, considering the sinc-type time autocorrelation function of the band-limited white noise. Accordingly, for a noise bandwidth of $\Delta f$ Hz, a maximum of $N=2 T_b \Delta f$ noise samples can be taken for each bit duration to ensure that the samples are statistically independent, which is also consistent with the Nyquist's sampling theorem. In other words, in accordance with the sampling theorem, faster-than-Nyquist sampling would result in correlated noise samples. Consequently, to increase $N$ to improve the accuracy of noise variance estimates, one needs to increase $T_b$; however, this reduces the bit rate given by $R_b=1/T_b$. This provides an interesting trade-off between the bit error rate (BER) and data rate for the KLJN system. Another attractive feature of KLJN is that no matter how strong Eve's signal processing capabilities are, the same limit also applies to her; she cannot gain further knowledge by expanding her observation interval or increasing the number of noise samples.
	
	\begin{figure}[!t]
		\begin{center}
			\includegraphics[width=0.65\columnwidth]{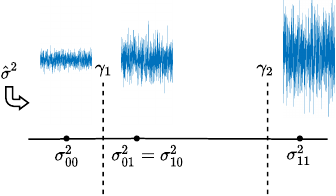}
			\vspace*{-0.3cm}\caption{Threshold-based noise voltage variance detection for the KLJN scheme (Reproduced from \cite{Basar_2023} with permission).}\vspace*{-0.3cm}
			\label{fig:Thresholds}
		\end{center}
	\end{figure}
	
	Let us denote the $k$th noise voltage sample by $x_k$, which is a Gaussian distributed random variable and observed simultaneously by Alice and Bob thanks to their Kirchhoff loop, that is, $x_k \sim \mathcal{N} (0, \sigma_i^2)$, where $\sigma_i^2 \in \left\lbrace 
	\sigma_{00}^2,\sigma_{01}^2,\sigma_{11}^2  \right\rbrace $ is the noise variance. Since the information is embedded in the noise variance, Alice and Bob process their samples to estimate it, as follows
	\begin{equation}
		\hat{\sigma}^2= \frac{1}{N} \sum\nolimits_{k=1}^N x_k^2.
	\end{equation}  
	An effective way for Alice and Bob to detect their partner's bit is to employ a threshold-based detector, as shown in Fig. 2, by employing two threshold values, $\gamma_1$ and $\gamma_2$. At this point, considering the value of $ \hat{\sigma}^2 $ as well as their own bit, Alice and Bob can determine their partner’s bit. For instance, assume that Alice's bit is $0$. In this case, she considers the first threshold, $\gamma_1$, to detect Bob's bit. If $ \hat{\sigma}^2 < \gamma$, her decision becomes $0$, and she marks the corresponding bit as an insecure one since she assumes that the selected bit combination is $00$. Particularly, she is aware of the fact that a third party can easily identify the case of $00$ through measurements on the line.
	
	On the contrary, if $ \hat{\sigma}^2 > \gamma$, her decision becomes $1$, which corresponds to an unconditionally secure bit exchange (generation). Since Alice and Bob's bits are different for this secure case, one can flip his/her bits to create a shared and secure key after the transmission. It is worth noting that for the detection scheme of Fig. 2, the error events between $00$ and $01/10$ dominate the overall BEP due to the proximity of the corresponding variance values on the decision space. Employing joint or alternating noise current samples, which provide an opposite relationship between noise variances compared to those given by voltage samples, this problem can be alleviated. As a result, much lower BEP can be achieved using more sophisticated detectors.
	
	It can be shown that the BEP of the KLJN scheme exponentially decays with the number of noise samples per bit, that is, $P_b\propto \exp(-N)$, which is promising to obtain very low BEP values for secure key generation between two partners. Furthermore, lowering the BEP does not compromise the unconditional security of the system. A detailed theoretical framework is represented in \cite{Basar_2023} by considering decision errors due to the randomness of the estimated noise variance. Furthermore, two new detectors are also presented in this study by exploiting joint voltage and noise samples further to improve the BEP performance of the KLJN scheme.

	\section{In Pursuit of Unconditional Security}
	One of the objectives during the introduction of KLJN in the early 2000s had been to challenge the following status quo on unconditional security: \textit{Only quantum physics can provide unconditional security for the key distribution}. At this point, the conceptual difference between post-quantum cryptography and quantum cryptography is worth noting. Specifically, post-quantum cryptography, also known as quantum-resistant or quantum-safe cryptography, deals with quantum-resistant cryptographic algorithms, which are implemented on existing platforms and derive their security through mathematical complexity. In principle, these algorithms will be resilient to attacks from quantum computers thanks to their highly complicated mathematical frameworks \cite{NSA_2020}. On the other hand, quantum cryptography, specifically quantum key distribution (QKD), uses principles of quantum mechanics to distribute the keys to be used later in a suitable form of cryptography and is genuinely unbreakable, unlike mathematical encryption. In other words, the security level provided by QKD is beyond the attackers of the future who possess quantum computers and even have infinite computational power\footnote{This strong assertion holds when we employ one-time pad ciphers with keys established using QKD, which will be discussed subsequently.}. In what follows, we dive into a historical journey on secure communication to catch valuable insights on the motivation of these physics-based key generation systems, KLJN and QKD, and their limitations in practice.
	
	The desire to communicate secretly has existed as long as writing and encrypted communication methods were developed in many ancient societies \cite{Sergienko_2006}. Notable historical examples include the scytale of Spartans from 400 BC, which performs a simple and yet practical transposition cipher, Caesar cipher from 50 BC, which performs a simple letter substitution method by replacing each letter in the message with the one appearing alphabetically three places apart, polyalphabetic ciphers in the 15th century, which are based on superpositions of Caesar ciphers with different shifts, and the famous Enigma machine of Arthur Scherbius developed after World War I, which uses electromechanical rotor technology to generate an encrypting sequence with an extremely long period of substitutions. Long story short, history is full of battles between cryptographers (codemakers) and cryptanalysts (codebreakers), with either side taking the lead at different points in time \cite{Singh_2000}. Considering the rise of physics-based and unconditionally secure key distribution systems, one can ask: Is it game over for codebreakers?
	
	After this quick introduction, this section presents the key distribution and keyless encryption approaches and their practical challenges. Finally, we briefly discuss the potential of PHY security approaches.
	
	\begin{figure}[!t]
		\begin{center}
			\includegraphics[width=1\columnwidth]{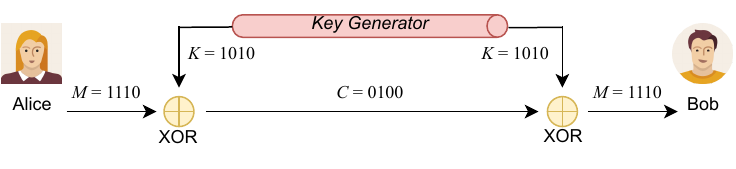}
			\vspace*{-0.4cm}\caption{One-time pad scheme. $M=1110 $: plaintext, $ K=1010 $: the shared secret key, and $ C=0100 $: ciphertext.}\vspace*{-0.3cm}
			\label{fig:OTP}
		\end{center}
	\end{figure}
	
	\subsection{Key Distribution Problem}
	
	On the search for truly unbreakable ciphers, the one-time pad (or the Vernam cipher \cite{Vernam}), illustrated in Fig. 3, shines out as a potential candidate to secure communications, once and for all. It was developed in the early 20th century and mathematically proven unbreakable by Shannon in the late 1940s \cite{Shannon_1949}. However, to ensure an unbreakable cipher using the one-time pad, the encryption key must be as long as the plaintext to be encrypted, it must be completely random, it must never be reused in whole or in part, and it must be kept completely secret between communicating parties. In other words, according to the terminology of Shannon, the entropy of the key ($ K $) must be higher than that of the message ($ M $). This brings out the \textit{secure key distribution problem}.

	The primary issue in secure communication is the key distribution problem \cite{Singh_2000,Djordjevic_2019}. A naive solution would be distributing the keys physically, as done by certain banks in the 1970s using trusted employees to convey decryption keys to their customers; however, this is not technically and logistically feasible by today’s standards. No matter how secure a cipher is in theory, it is challenged by key distribution problems in practice. In other words, if two parties want to communicate securely, they must rely on a third party to deliver the key, which is the Achilles heel of the system.
	
	Interestingly, the entire key distribution problem embodies a catch-22 situation, a paradoxical scenario one cannot escape due to its contradictory rules or limitations. When two individuals aim to exchange a secret message, the sender must first encrypt it. However, for encrypting the secret message, the sender must utilize a key that remains a secret. This dilemma presents the challenge of transmitting the secret key to the receiver to enable the transmission of the encrypted secret message. Consequently, to effectively communicate a secret message, the two users must already share a secret (the key).
	
	Despite the most revolutionary development in 20th-century cryptography being the creation of effective mathematical techniques to overcome this problem, the extensive implementation of quantum computers in the future might challenge their computational (conditional) security. At this juncture, a pertinent question arises: Can a secret message be securely communicated without the need for exchanging a key?

	\subsection{Keyless Encryption and Practical Issues}
	
	\begin{figure}[!t]
		\begin{center}
			\includegraphics[width=1\columnwidth]{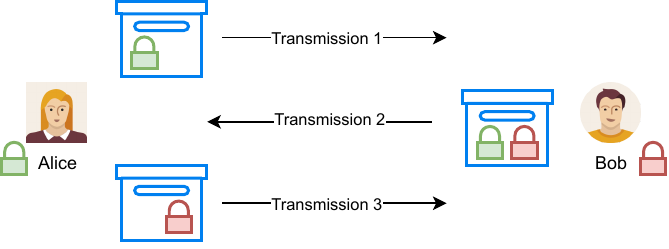}
			\vspace*{-0.3cm}\caption{Three-pass protocol realized by double padlocks and a box containing the secure message.}\vspace*{-0.3cm}
			\label{fig:THREE_PASS}
		\end{center}
	\end{figure}
	
	For 2000 years, the distribution of keys has been regarded as an indisputable aspect of secure communication. However, challenging this dogma, the doubly encrypted scheme (Shamir’s three-pass or no-key protocol developed in the 1970s) enables the exchange of secure messages without the need to exchange a key \cite{Massey_1988}. For the first time, we think that key exchange might not be an inevitable part of secure communication! The three-pass protocol can be better understood by invoking the well-known mechanical analogy of a mail carrier box with two padlocks, as illustrated in Fig. 4 \cite{Kish_2004,Chappell_2014}. Assume that Alice intends to send a highly confidential message to Bob, accounting for an unreliable carrier service. She places the message in a box, secures it with her green padlock (operation $ A $), and then dispatches it to Bob (Stage 1). Upon receiving the box, Bob cannot unveil the secret message as he lacks the key to Alice’s padlock. Conversely, he adds his red padlock (operation $ B $) and returns the double padlocked box to Alice (Stage 2). At this point, Alice removes her green padlock and forwards the box back to Bob (Stage 3). The returned box only contains Bob’s padlock, which he removes to access the secret message. Thus, secure transmission has been achieved in three stages, all without a key involvement!

	It appears that the problem of key distribution is resolved, yet several fundamental challenges remain to be tackled. These challenges encompass the development of practical mathematical tools (involving classical physical operators) that incorporate randomness and adhere to the specific sequence of Alice and Bob’s locking/unlocking operations, ensuring an unconditionally secure and implementable three-pass scheme. In this context, it is necessary to ensure that even if Eve, who represents one of the central components of this unreliable carrier service between Alice and Bob, observes all three publicly shared messages, she cannot recover the concealed message. Additionally,  it is crucial that Alice and Bob share no information about their actual operators $ A $ and $ B $.

	The practical problems associated with the three-pass protocol have resulted in the development of the Diffie-Hellman-Merkle (DHM) key exchange scheme in 1976. This scheme involves modular arithmetic $ (Y^x (\mathrm{mod}\, P)) $  to facilitate key exchange between Alice and Bob through public discussion \cite{Singh_2000}. Nevertheless, the three-pass protocol and the DHM key exchange scheme share a significant drawback: Alice and Bob must exchange messages and wait for each other's responses. The invention of public key cryptography (also known as asymmetric cryptography) by Rivest, Shamir, and Adleman in 1977 not only addressed this waiting problem stemming from public discussions but also overcame the challenge of key distribution by leveraging public and private keys, which are based on huge prime numbers. To illustrate with a mechanical analogy, Alice creates numerous replicas of her padlock and distributes them to various mail offices in this scenario. A sender wishing to convey a secret message to Alice places it in a box and employs a replica of Alice’s padlock (public key) before sending it. Only Alice can unlock this box, as she is the sole possessor of the private key for these padlocks. In essence, anyone can secure a padlock by simply snapping it shut; however, only the person with the private key can open it. A notable drawback of asymmetric encryption is its demand for more time and computing power, which may not be suited for devices with limited memory and computational resources. On the other hand, asymmetric encryption can serve as a secure method for exchanging a symmetric key to be utilized in bulk encryption on suitable devices \cite{Sattler_2022}.

	\subsection{Release the Kraken: Security at the PHY Level}
	
	The high computing power and processing time requirements of modern encryption algorithms have been the major driving force behind the emergence of PHY security solutions that have flourished in the past decade. Specifically, PHY security provides node authenticity, data confidentiality, and data integrity by leveraging the dynamic characteristics of the wireless environment \cite{Shakiba-Herfeh_2021,Zhou_2016,Poor_2016}. Despite the solid security solutions that PHY security offers, considering it as a quantum-resistant solution would be a naive assumption. This can be attributed to the fact that PHY security generally relies on optimistic assumptions concerning the availability of perfect channel state information (including that of the eavesdropper's), multiple antennas, rich scattering environments, rapidly varying channels, infinite transmission power, and better channel conditions of legitimate receivers \cite{Solaija_2022}. In other words, despite its solid theoretical appeal, PHY security often achieves only conditional security in many practical scenarios, meaning that Alice and Bob must have an advantage over Eve within the PHY itself. This limitation is one of the factors hindering its widespread use in current networks.

	Secret-key generation emerges as one of the promising applications of PHY security. In this context, Alice and Bob can leverage the wireless channel to generate correlated random sequences and subsequently use a public authenticated and error-free feedback channel to agree upon a secret key \cite{Zeng_2015}. In the source-type model, terminals observe the correlated output of the randomness source to generate keys; however, challenges remain in implementing this in practice \cite{Djordjevic_2019}. In the channel-type model, one terminal transmits random symbols to other terminals using a broadcast channel, which opens up new opportunities. Nevertheless, the security of both models can be compromised under active attacks, such as jamming and channel manipulation. Moreover, most of the available PHY key generation schemes require a noiseless public discussion channel with unlimited capacity.

\begin{figure}[!t]
		\begin{center}
			\includegraphics[width=1\columnwidth]{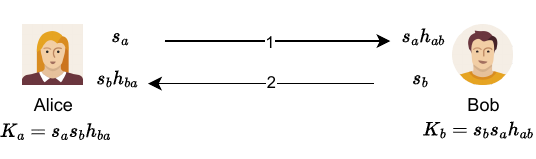}
			\vspace*{-0.3cm}\caption{Secret key generation with random signals of Alice ($ s_a $) and Bob ($ s_b $) and reciprocal channel ($ h_{ab}=h_{ba} $).}\vspace*{-0.3cm}
			\label{fig:Key_generation}
		\end{center}
	\end{figure}

	In light of the above discussion, secret key generation with user-generated randomness, illustrated in Fig. \ref{fig:Key_generation}, stands out as a promising direction \cite{Zeng_2015}. Here, using two transmission intervals, a shared secret key ($K_a=K_b$) is generated considering both user-generated and channel-oriented randomness and channel reciprocity. In simple terms, Alice and Bob multiply their received signals carrying the random messages of their partner by the locally generated random signals to compose a shared randomness. It is worth emphasizing the conceptual similarity between this solution and the DHM key exchange scheme.
	
	\begin{figure*}[!t]
		\begin{center}
			\includegraphics[width=2\columnwidth]{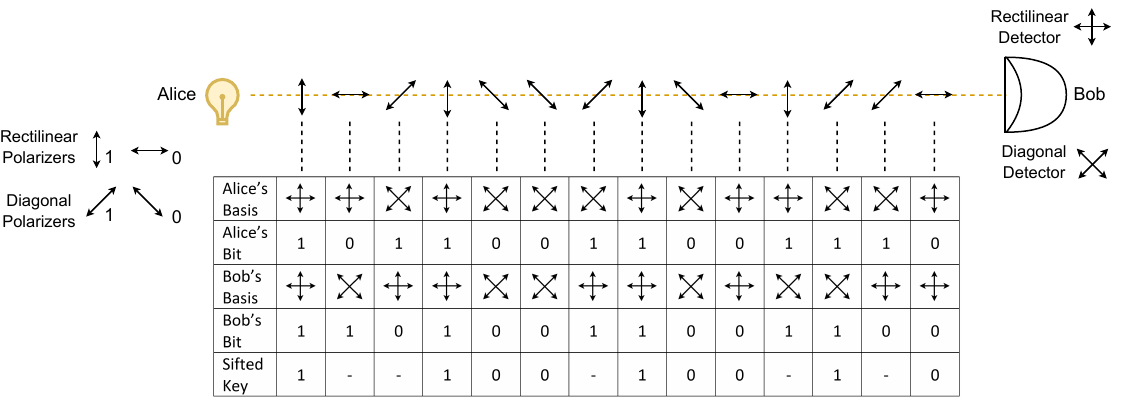}
			\vspace*{-0.1cm}\caption{QKD in action: BB84 protocol.}\vspace*{-0.3cm}
			\label{fig:QKD}
		\end{center}
	\end{figure*}

	The use of AN-based systems introduces another new dimension to PHY security. In this approach, the transmitter allocates a portion of its available power to generate AN signals, strategically degrading only the eavesdropper's channel and thereby ensuring communication security \cite{Goel_2008,He_2017}. Despite its significant potential, most of the AN-based systems in the literature utilize MIMO antennas to nullify the impact of AN on the legitimate receiver, essentially diminishing the signal-to-noise ratio (SNR) experienced by Eve \cite{Guan_2020}. In essence, the primary goal of AN-based systems is to reduce Eve's SNR, not to use the AN itself for secrecy. However, assuring that Eve's BER approaches $ 0.5 $ when the AN power is constrained is challenging, as some confidential information might inadvertently still reach her. These challenges contribute to the conditional security nature of AN-based systems.

	In conclusion, while public key cryptography is widely used, it offers only conditional (computational) security and could be compromised by quantum computers within a feasible timeframe. PHY security is a complementary solution to address the computational demands and latency constraints associated with higher-level cryptographic techniques. However, it is important to note that the security of PHY solutions relies on external factors. Given these considerations, our attention naturally shifts towards physics-based key distribution systems like KLJN and QKD as we seek to attain the pinnacle of unconditional security.

	\section{Quantum Key Distribution: Game Over for Codebreakers?}
	QKD addresses one of society's most pressing concerns: unconditionally secure communication. Specifically, QKD systems leverage the principles of quantum mechanics to create a secret key between two distant users. More importantly, the security of this key is guaranteed by the laws of physics \cite{Djordjevic_2019,Wolf_2021}. A unique property of QKD is its users’ ability to detect the presence of a third party in between, either passive or active since measuring a quantum system disturbs it. In other words, an eavesdropper trying to gain access to the key must perform measurements, which cause detectable anomalies at QKD transmitter and receiver nodes. Lastly, it is noteworthy that while QKD cannot be used for secure data transmission, similar to KLJN, it can generate secure keys for subsequent bulk data encryption, i.e., it is a key establishment protocol.

	The following example can clearly illustrate the operation of QKD: BB84 protocol developed by Bennett and Brassard in 1984 \cite{Bennett_2014}. First, Alice randomly selects one of two available bases (alphabets): rectilinear basis (\ooalign{$\leftrightarrow$\cr\hfil$\updownarrow$\hfil}) or diagonal basis (\ooalign{$\neswarrow$\cr\hfil$\nwsearrow$\hfil}) for the polarization of her photon. In the rectilinear basis, the photon is either vertically ($ \updownarrow $) or horizontally ($\leftrightarrow$) polarized. In the diagonal basis, the polarization is either $+45^{\circ}$ ($ \neswarrow $) or $-45^{\circ}$ ($\nwsearrow $). After the basis is selected, a classical bit ($ 0 $ or $ 1 $) can be encoded to the polarization, as shown in Fig. 6. For instance, we can assign Bit $ 1 $:$ \updownarrow $ and Bit $ 0 $:$\leftrightarrow$ for Basis \ooalign{$\leftrightarrow$\cr\hfil$\updownarrow$\hfil} and Bit $ 1 $:$ \neswarrow $ and Bit $ 0 $:$\nwsearrow $ for Basis \ooalign{$\neswarrow$\cr\hfil$\nwsearrow$\hfil}. According to quantum mechanics, Bob’s detection can be correct if and only if he selects the same basis as Alice; otherwise, he ends up with a wrong bit interpretation. Specifically, suppose a diagonally polarized photon is measured in the rectilinear basis (or vice versa). In that case, a completely random polarization result is obtained, meaning no information cannot be gained. According to the principles of quantum physics, the state of the target photon is changed irreversibly in this measurement process. As a result, it is not possible to determine its polarization before passing the corresponding detector at the QKD receiver.
	
	Since Bob does not know the bases Alice encoded in the photons, he can only select a basis at random to measure the incoming photon, either rectilinear or diagonal bases (rectilinear or diagonal detectors). After this step, Alice and Bob hold a bit sequence called the raw key pair. At this point, they rely on their public discussions to filter out this raw key pair. Through a public channel, Bob announces the bases he has randomly selected to measure Alice's photons. After comparing Bob's bases to the ones she used in photon generation, Alice responds to Bob by stating the cases in their selections coincide. Finally, Alice and Bob discard the corresponding bits for which the encoding and measurement bases are not the same, and they obtain the shared secret (sifted) key.
	
	Let us give an example considering the QKD scheme given in Fig. 6. For the first transmission, Alice and Bob's bases coincide (both rectilinear); as a result, Bob's detection of bit $1$ is correct. On the contrary, for the second transmission, Bob's randomly selected basis (diagonal) is different from that of Alice's (rectilinear), and the bit interpretation of Bob is inconclusive. Specifically, Bob can obtain either bit $0$ or bit $1$ with $50\%$ probability after detection with the wrong basis, and he discards this bit in the sifted key after public discussion of selected bases with Alice. For the transmission given in Fig. 6 with fourteen bits, the bases of Alice and Bob coincide for photons $1 , 4, 5, 6, 8, 9, 10, 12, $ and $ 14 $, and the shared key is obtained as $ 110010010 $ after public discussion of selected bases.
	
	After 2000 years, is it game over for codebreakers? At different times in history, cryptographers have believed that the monoalphabetic/polyalphabetic ciphers and mechanical ciphers such as Enigma were all unbreakable; however, their claims were eventually proved wrong. This can be explained by the fact that all these ciphers were developed, assuming that they were complex enough to challenge the ingenuity and the technology of codebreakers at that time. On the other hand, quantum cryptography exploiting QKD is totally unbreakable under ideal conditions, and this claim is fundamentally different from the earlier ones. This is one of the reasons why codebreakers are transforming into physicists and engineers to understand and exploit the practical imperfections of QKD (and KLJN) systems to gain knowledge on secure keys. Here, we can ask the following questions: What are the significant disadvantages of QKD in practice? Why is post-quantum cryptography favored more than QKD for practical networks, for instance, by authorities \cite{NSA_2020}? These exciting questions remain to be answered.

	\section{Clash Of the Titans: KLJN vs QKD}
	
	The KLJN scheme stands out as a potential alternative to QKD, using statistical-physical techniques for the secure distribution of shared randomness. Nevertheless, despite our fancy title, comparing KLJN and QKD is the comparison of apples and oranges since both schemes have entirely different architectures and characteristics. Furthermore, both designs have pros and cons and might suit different applications and users. For instance, KLJN has a much lower deployment cost and might be more suitable for low-rate ($\sim\! 100 $ bit/sec) key distribution applications at moderate distances ($ 2-20 $ km) or even shorter ranges, such as inside data centers and computers \cite{Mingesz_2008}. Specifically, a bandwidth-distance product of $ 
	2\times 10^6 $ Hz-meter is reported for the KLJN scheme due to the distance limitations imposed by the Kirchhoff law \cite{Kish_2006}. On the other hand, QKD systems have a significantly higher cost ($\sim\$100$K) due to their sophisticated hardware, however, off-the-shelf QKD products\footnote{A notable commercial off-the-shelf product is the 4th generation Clavis XG QKD system from ID Quantique developed in May 2022 \cite{Clavis}.} can support data rates of more than $ 100  $ kbits/sec over distances higher than $ 100 $ km. As a result, instead of putting them into competition, we have to assess their potential for specific use cases. In this sense, this article aims not to enter the ongoing debates between QKD and KLJN supporters from the past decade. Instead, it aims to shed light on their operating principles and unconditional security aspects with their potential pros and cons.
	
	\subsection{A Brief Review of KLJN Literature}
	
	Despite its undeniable potential, the literature on KLJN is relatively limited due to its multidisciplinary nature. In this context, most of the studies in the literature focus on different types of KLJN attacks and corresponding defense mechanisms. A limited number of enhanced KLJN designs and performance analyses have also been reported recently.
	
	The real-time performance of the KLJN system is investigated by practical experiments at several distances and data rates for the first time in \cite{Mingesz_2008} using external noise generators. In this study, a noticeable BER of $ 0.02\% $ is obtained through practical experiments. Wire resistance effect on the noise voltage and current, which might be inevitable in practical setups, is investigated in \cite{Kish_Scheuer_2010}. Here, the information leak to Eve stemming from non-zero wire resistance is reported to be not significant. It is worth noting that Gaussian-distributed noise signals (either natural background or externally generated ones) must be used in KLJN systems for the secure key exchange, as mathematically demonstrated in \cite{Gingl_2014}. Furthermore, the use of artificial noise generators in practical setups allows for the generation of strong signals (noise temperatures corresponding beyond a trillion Kelvin) and the control of temperature with astronomically high accuracy \cite{Kish_2016}.
	
	An inspiring generalized KLJN scheme is introduced in \cite{Vadai_2015} using arbitrary (four different) resistors. However, compared to the original KLJN scheme, this generalized scheme is shown to be less secure in the presence of passive attacks utilizing the different line resistances for $01$ and $10$ cases \cite{Ferdous_2020}. The theoretical BEP performance of the KLJN scheme is investigated for the first time in \cite{Saez_2013}, and an exponentially decaying error probability is obtained concerning the bit duration, which might be very promising for reliable bit exchange. The joint use of voltage and current samples (measurements) is considered in the follow-up study of \cite{Saez_2013b} to further reduce the BEP of KLJN systems by using more sophisticated protocols. However, both of these studies considered only the performance of securely exchanged bits, assuming that the protocol discards the insecurely detected ones and certain theoretical approximations are used to derive the theoretical BEP. In \cite{Smulko_2014}, using statistical hypothesis testing, a new and systematic approach is used to assess the effect of relative noise variance ratios on the BEP performance of the KLJN system. Promising analog KLJN variants such as random-resistor KLJN and random-resistor random-temperature KLJN are introduced to increase the key rate and improve the immunity against earlier attacks \cite{Kish_Granqvist_2016}. Recently, a comprehensive framework has been presented in \cite{Basar_2023} on the BEP optimization and the receiver system design of the KLJN scheme. In this study, not only guidelines are presented for the KLJN system designer but also two new KLJN detectors utilizing joint and alternating voltage and current samples are proposed to further improve its BEP.
	
In Fig. 7, we present the timeline of some major KLJN developments and the corresponding studies from the literature reviewed in this subsection. 

\begin{figure}[!t]
	\begin{center}
		\includegraphics[width=0.75\columnwidth]{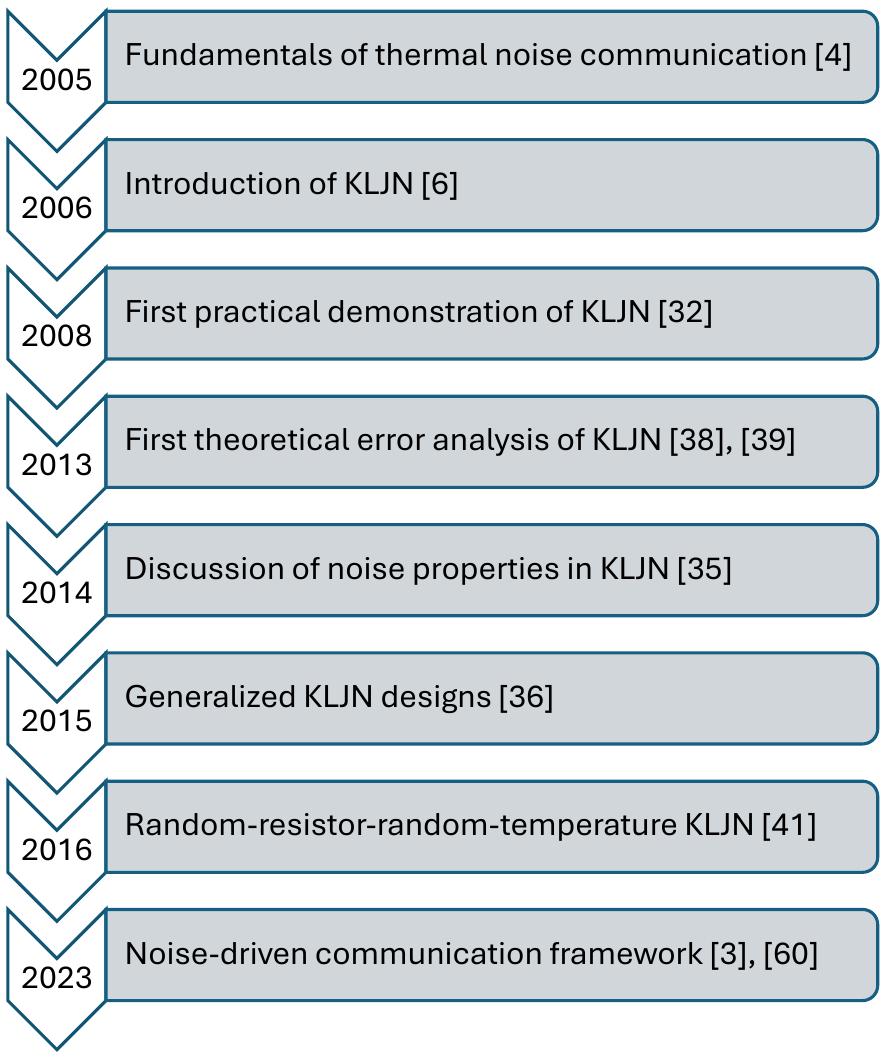}
		\vspace*{-0.3cm}\caption{Timeline of some major KLJN developments.}\vspace*{-0.3cm}
		\label{fig:KLJN_timeline}
	\end{center}
\end{figure}

	A variety of attack types have been proposed for the KLJN scheme in the past decade: passive time-correlation attacks \cite{Bennett_2013}, directional wave measurement attacks \cite{Gunn_2014}, transient wave attacks \cite{Gunn_2015}, cable resistance and capacitance based attacks \cite{Kish_Granqvist_2014,Chen_2015}, current injection attacks \cite{Chen_2016}, random number generation attacks \cite{Chamon_2021}, and many more. We note that some of these attacks are optimistic and might not work in practice, while others might work but have efficient defence mechanisms. While a detailed coverage of these important attacks, as well as their defense mechanisms, are beyond the scope of this review article, interested readers are referred to \cite{Kish_2016} and the references therein for a detailed investigation of these.
	
	Through the authenticated exchange of Alice and Bob’s current and voltage measurement data, a general defense method can be provided against these active (invasive) attacks, which inject or extract energy from the line \cite{Chen_2016,Kish_2006_New}. Comparing the currents and voltages at the two ends also eliminates the man-in-the-middle (MITM) attack for KLJN; however, this requires an authenticated and classical communication channel between Alice and Bob. In this context, QKD and KLJN schemes show similarities in their operation to maintain their integrity against specific attacks.
	
				\begin{table*}[t]
		\centering
		\caption{Comparison of certain KPIs of two physics-based key distribution systems: KLJN and QKD.}\hspace*{-0.5cm}
		{\footnotesize  \begin{tabular}{l|l|l|l|l|l|l}
				& Key Rate                            & Range                              & Complexity                         & Cost                         & BER                          & Major Protocols                     \\ \hline
				KLJN & \cellcolor[HTML]{FD6864}Low         & \cellcolor[HTML]{FFCC67}Low-Medium & \cellcolor[HTML]{9AFF99}Low-Medium & \cellcolor[HTML]{9AFF99}Low  & \cellcolor[HTML]{9AFF99}Low  & Classical, Intelligent, Analog KLJN \\ \hline
				QKD  & \cellcolor[HTML]{9AFF99}Medium-High & \cellcolor[HTML]{9AFF99}High       & \cellcolor[HTML]{FD6864}High       & \cellcolor[HTML]{FD6864}High & \cellcolor[HTML]{FD6864}High & BB84, B92, SARG04, $6$-state        \\ \hline
		\end{tabular}}
	\end{table*}
	
	\subsection{Background of QKD Protocols}
	
	On the contrary, academic, commercial, and governmental interest in QKD systems has flourished remarkably in the past decade. QKD systems have rapidly expanded in recent years, with numerous new commercial products and terrestrial networks being established. In the following, we present a concise overview of the QKD literature, primarily emphasizing communication engineering and algorithmic design aspects.
	
	The first QKD protocol was proposed by Bennett \& Brassard in 1984, as briefly discussed earlier \cite{Bennett_2014}. Specifically, different properties of photons, such as time, frequency, phase, and orbital angular momentum, can be employed to implement various QKD protocols along with continuous-variable QKD systems. Most commercial QKD products are also based on the popular BB84 protocol, a discrete-variable QKD system built on the no-cloning theorem of quantum physics and the irreversibility of measured quantum states. Due to the possibility of an MITM attack by Eve, the sifting phase must be authenticated by Alice and Bob. This phase generally consists of three messages between Alice and Bob. Two messages are sent by Bob, where he tells Alice the photons he was able to measure and the bases he used for the measurement of those. Alice completes the sifting phase by telling Bob the bases she used for those photons measured by Bob. However, this bit sifting with authentication requires a classical communication channel, as any unconditionally secure system.
	
	B92 protocol developed by Bennett in 1992 uses only two non-orthogonal states in Alice to represent bit $ 0 $ and bit $ 1 $ \cite{Bennett_1992}. An important contrast to the BB84 protocol is that Bob does not have to announce the choice of his bases in the post-transmission communication. In other words, bit sifting is not required in BB92 due to its clever polarization selection algorithm. The SARG04 protocol uses the same four states and the same measurements on Bob’s side as BB84, but the bit is coded in the basis rather than in the state \cite{Scarani_2004}. The advantage of this protocol is that Alice does not need to reveal her choice of basis.
	
	After all these three protocols, Alice and Bob perform classical post-processing, including information reconciliation and privacy ampliﬁcation, to remove the errors stemming from the channel and to reduce the correlation that remains with Eve, respectively. An overview of the rich set of QKD protocols can be found in \cite{Cao_2022}.

	In terms of the employed QKD system types, the BB84 protocol belongs to device-dependent QKD, while another QKD system type is measurement-device-independent QKD (MDI-QKD) \cite{Djordjevic_2019}. Here, a third node (Charlie), which employs quantum detectors, is involved, while Alice and Bob have quantum sources. After preparing their quantum states, Alice and Bob transmit these to Charlie's detectors. By performing partial Bell state measurements, Charlie raises a flag when the desired partial Bell states are detected. Using a third node inevitably introduces new opportunities for QKD but also renders the overall system more complicated.

	Two prevalent QKD attacks are intercept-resend (IR) and photon number splitting (PNS) attacks \cite{Brassard_2000,Djordjevic_2019}. In the former, Eve intercepts the quantum signal sent by Alice and measures it. Based on the measurement result, she prepares a new quantum signal in the same measured state to be sent to Bob. Due to quantum weirdness, Eve can only detect correct states with a probability of $ 1/2 $. In return, the overall error probability between Alice and Bob becomes $ 1/4 $, which is relatively high and can be easily noticed by them. However, in the more sophisticated PNS attack, Eve exploits the fact that Alice transmits multiple photons for each bit. Specifically, Alice’s coherent source emits photons according to the Poisson distribution with a typical average photon number of $ 0.1-0.8  $ (Poisson parameter $\alpha$) for weak laser pulses \cite{Wolf_2021}. Despite its rare occurrence probability, in this case, Eve employs a beamsplitter to take one of the photons and pass the rest to Bob. To circumvent this, Alice can employ a decoy-state-based protocol. In this protocol, to represent the signal and decoy states, Alice transmits quantum states with different mean photon numbers to confuse Eve \cite{Lo_2005}. While the literature contains a rich set of other attacks involving heavy quantum physics and numerous debates about the practical vulnerabilities of QKD systems, we conclude our discussion at this point.

	\subsection{KLJN vs QKD}
	
	Table I presents a high-level comparison of KLJN and QKD systems considering their major key performance indicators (KPIs), such as key rate, range, complexity, cost, and BER. Here, we use high, medium, and low labels based on specific KPIs for each system, and red and green colors are respectively used for negative and positive aspects of each system for the corresponding KPI. For instance, QKD outperforms KLJN regarding key rate, benefiting from not requiring a circuit loop between two users. Nevertheless, sophisticated attacks utilizing photon groups can limit the key rate of practical QKD systems. However, QKD and KLJN systems have different power scaling laws and the target range is critical in their comparison. We note that the complexity of KLJN can be as high as QKD when fully defended but its cost can be very low  if integrated on a chip. In the same table, we also list the corresponding major protocols associated with these two physics-based key distribution systems. We conclude that both designs can offer unconditional security by exploiting the laws of physics under certain conditions. However, the choice between these two key distribution approaches should be made carefully, depending on the requirements of the target applications. Evaluating the performance of two concepts through computer simulations and practical experiments is also a promising future research direction. At this point, we also pose the following interesting question: can these two different key distribution concepts merged to create an ultimate key distribution system?
	
Finally, it is worth noting that both KLJN and QKD systems are affected by real-world conditions and can be used cautiously by system designers in the presence of certain effects. For instance, the thermal equilibrium of the two resistors of the KLJN system should be preserved to guarantee perfect security and eliminate cable resistance-based attacks \cite{Kish_2016}. Furthermore, in practical setups, wire resistance should be much less than Alice and Bob’s resistances, and the wire capacitance should not affect the loop impedance. Proper filtering and advanced detectors might also be necessary to reach the theoretical key rates of KLJN systems and to overcome the deviations caused by environmental factors. Furthermore, there might be an information leak in non-ideal practical setups, while unconditional security can still be preserved by various techniques such as privacy amplification. On the other hand, practical QKD systems also suffer from short range, low bit rate, and high cost. Furthermore, generating strictly ideal single photon packages and avoiding detector noise is impossible for QKD systems under practical constraints.

To conclude, both KLJN and QKD systems rely on specific laws of physics to create secure keys; however, they might require additional defense mechanisms in practice to preserve unconditional security. If implemented at the chip level, the low cost of KLJN systems makes them a potential candidate for short-range critical applications in which system designers can create effective Kirchhoff loops. On the other hand, QKD systems require dedicated hardware, have a much higher cost, and can be suitable for more advanced and long-range key distribution applications. 

			\begin{figure*}[!t]
		\begin{center}
			\includegraphics[width=1.7\columnwidth]{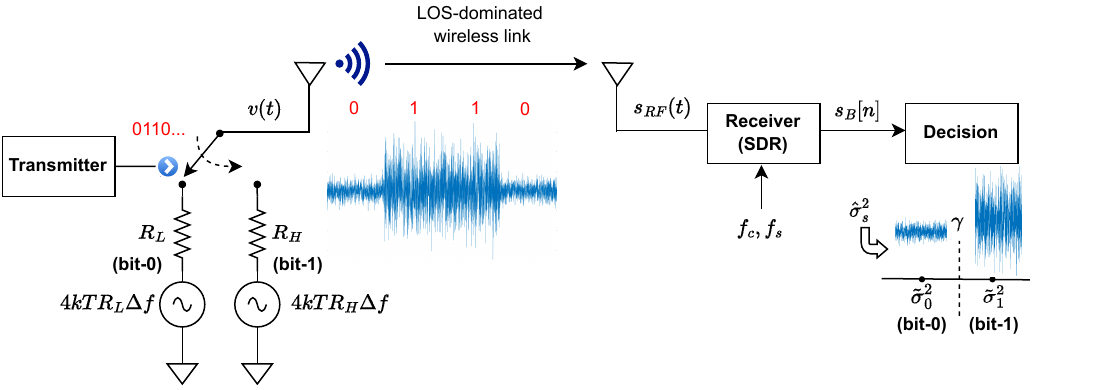}
			\vspace*{-0.1cm}\caption{Wireless thermal noise modulation. Two representative thermal noise voltage $(v(t))$ variances are shown for bit $0$ and bit $1$ cases ($0\rightarrow$ low and $1\rightarrow$ high). $s_{RF}(t)/s_B[n]$: received passband/baseband signals, $f_c/f_s$: center/sampling frequency, $\hat{\sigma}_s^2$: estimated noise variance, $\hat{\sigma}_0^2/\hat{\sigma}_1^2$: expected complex baseband noise variances for bit $0/1$, and $\gamma$: comparison threshold. (Reproduced from \cite{Basar_2023} with permission). }\vspace*{-0.3cm}
			\label{fig:TherMod}
		\end{center}
	\end{figure*}

	\section{Noise is in the Air}
	Breaking the Kirchhoff loop of KLJN and performing IM for resistances over wireless channels has led wireless \textit{thermal noise modulation (TherMod)}, which is illustrated in Fig. 8. In this setup, incoming information bits at the transmitter determine the index of the selected resistor similar to the KLJN scheme. This selection allows the modulation of the power spectral density of the generated thermal noise waveform, as initially demonstrated in the early study of \cite{Kish_2005}. Despite the conceptual similarity between KLJN and TherMod, the latter cannot provide unconditionally secure communication due to the non-existence of a noise loop between transmitter and receiver. However, if implemented by background noises, a transmitter with extremely low power consumption (less than $1$ microwatt as backscatter communication (BackCom) systems) can be obtained, which also provides stealth communication capabilities. Another advantage of TherMod is that its transmitter does not rely on existing RF signals to operate as ambient BackCom systems \cite{Liu_2019}.
	
	The concept of TherMod was envisioned in the early work of \cite{Kish_2005} by two parabolic antennas driven by variable resistances (open, short, and 50 ohms). An experimental demonstration of this scheme has been recently given in \cite{Kapetanovic_2021}, where due to extremely low thermal noise powers, a short range and line-of-sight (LOS)-dominated communication setup is considered with highly directional horn antennas. Nevertheless, by simply indexing resistances, the authors reported a maximum data rate of $26$ bits/second for short distances and a maximum range of $7.3$ meters. Although some early designs have been reported in the literature to convey information by random signal parameters \cite{Kramer_2017,Basnayaka_2017,Kozlenko_2018,Anand_2019}, such as variance, we have provided a general framework by using resistance IM to convey information and obtained comprehensive theoretical results depending on specific system parameters, such as useful and disruptive noise powers \cite{Basar_2023}. Recently, the noise modulation concept is also proposed to use externally generated noise signals and a number of noise-alike waveforms are created to convey information bits through wireless channels \cite{Basar_2023c}.
	
	Similar to the KLJN system, the task of the TherMod receiver is to extract the information embedded in the variance of the noise signals generated at the transmitter. For this purpose, the receiver software-defined radio (SDR) device can take samples from the received waveform and obtain complex baseband samples for further processing. Here, the level of the additional noise the receiver chain introduces should be as low as possible to make a reliable bit detection. For instance, as shown in Fig. 8, calculating the variance estimate $(\hat{\sigma}_s^2)$, 
	a decision on the transmitted bit can be made by comparing it with a suitable threshold.
	
	In contrast to resistor indexing, in the scheme of noise loop modulation \cite{Mucchi_2022}, by creating a full-duplex noise communication loop between the transmitter and the receiver, the noise itself is modulated to obtain a solid secrecy at the PHY level for both additive white Gaussian noise and multipath fading channels. In other words, this two-way loop ensures that there is always a mixture of the user bits and noises over the air, and a  third node with different noise processes cannot reveal the transmitted information, no matter its computational power. The scheme of noise loop modulation appears as a promising candidate to use the noise signals themselves for unconditional security over wireless channels, differently from artificial noise-based PHY security systems that mainly aim to degrade the signal-to-noise ratio of potential eavesdroppers \cite{He_2017}. On the downside, the data rate of this scheme is relatively limited, and it requires a full-duplex communication loop to operate. Furthermore, the effect of changing bit values at both sides (transients) should be carefully examined from the perspective of preserving the secrecy.

	\section{Conclusions and Future Directions}
	In this review article, under the central theme of \textit{Kirchhoff meets Johnson}, we have first introduced the fundamentals of thermal noise-driven communication and the KLJN scheme. Then, using the KLJN scheme as a reference point, we have presented a historical journey toward unconditionally secure communication and revisited the longstanding yet critical challenge in this field: secure key distribution. In pursuing unconditionally secure communication empowered by physics-based key distribution schemes, we also conducted a detailed review of QKD systems and provided concise comparisons with KLJN systems. Lastly, we have explored over-the-air thermal noise transmission concepts and evaluated their promising potential for low-power, stealth, and secure communications.
	
The major advantages and weaknesses of KLJN-based systems are summarized as follows. First, the KLJN systems can be deployed at a much lower cost if implemented at the chip level and can secure short- to medium-range links effectively without requiring significant maintenance. Utilizing more advanced detectors with joint noise voltage and current samples, the error probability of KLJN systems can be reduced further, which might help increase the range and key rate. As a result, KLJN-based solutions shine out for potential applications such as securing communications within data centers and local networks. However, the cost of the network will increase to defend the system against various attacks fully. The requirement for an active Kirchoff loop also hinders the flexibility of KLJN-based systems to operate at challenging conditions and long ranges.
	
Potential future research and development directions for unconditionally secure communication systems include the generalization of TherCom schemes for an arbitrary number of resistors and external sources, the exploration of practical issues such as wire resistances, sampling imperfections, and timing errors, the proposal of suitable coding schemes or effective bit mapping algorithms to improve the error performance further, and the integration of these concepts into legacy communication systems. Mainly, addressing the limited key distribution range and key rate of KLJN is a challenge that needs resolution for its widespread practical use. Furthermore, developing novel KLJN variants that can be used for two-way secure data transfer is an interesting research direction. In such a setup, KLJN communicators can be directly used to exchange secure data between two users, but the transmission needs to be unconditionally secure for all bit cases. QKD systems’ very high cost and fragile structure also require careful assessment for their practical applications. In this context, alternative key distribution solutions such as KLJN can be evaluated in certain cases to complement their quantum-based counterparts. At this point, merging these two key distribution concepts might also be an interesting problem in creating an ultimate unconditionally key distribution system. However, the security of both designs should be meticulously examined under practical imperfections and constraints. Furthermore, developing unconditionally secure PHY security signaling techniques empowered by noise-alike signals and providing a complete uncertainty for hidden eavesdroppers at the PHY layer is an exciting future research direction.
	
	We ask again: Is it game over for codebreakers after their relentless efforts to outwit codemakers? In other words, with the advent of unconditionally secure systems, has the millennia-old battle between codebreakers and codemakers finally concluded? While these complex queries do not yield quick answers, the emergence of physics-based secure key distribution techniques and their potential for widespread adoption soon imply that codebreakers would encounter a formidable challenge when attempting to breach unconditionally secure communication systems.

\bibliographystyle{IEEEtran}
\bibliography{IEEEabrv,bib_2023}


\begin{IEEEbiographynophoto}{Ertugrul Basar}(Fellow, IEEE) received his Ph.D. degree from Istanbul Technical University in 2013. He is currently an Associate Professor with the Department of Electrical and Electronics Engineering, Koç University, Istanbul, Turkey and the director of Communications Research and Innovation Laboratory (CoreLab). He had visiting positions at Ruhr University Bochum, Germany (2022, Mercator Fellow) and Princeton University, USA (2011-2012, Visiting Research Collaborator). His primary research interests include beyond 5G and 6G wireless systems, MIMO systems, index modulation, reconfigurable intelligent surfaces, waveform design, zero-power and thermal noise communications, physical layer security, quantum key distribution systems, and signal processing/deep learning for communications.
	
In the past, Dr. Basar served as an Editor/Senior Editor for many journals including \textsc{IEEE Communications Letters} (2016-2022), \textsc{IEEE Transactions on Communications} (2018-2022), \textit{Physical Communication} (2017-2020), and \textsc{IEEE Access} (2016-2018). Currently, he is an Editor of \textit{Frontiers in Communications and Networks}. He is an Associate Member of Turkish Academy of Sciences (2023). 
\end{IEEEbiographynophoto}

\end{document}